# Vanadium-Based Superconductivity in a Breathing Kagome Compound Ta$_2$V$_{3.1}$Si$_{0.9}$

HongXiong Liu[1,2#], JingYu Yao[1,3#], JianMin Shi[4#], ZhiLong Yang[1#], DaYu Yan[1], Yong Li[1,2], DaiHong Chen[4], Hai L. Feng[1,5], ShiLiang Li[1,3,5], ZhiJun Wang[1,3]*, YouGuo Shi[1,2,5]*

[1]*Beijing National Laboratory for Condensed Matter Physics and Institute of Physics, Chinese Academy of Sciences, Beijing 100190, China*

[2]*Center of Materials Science and Optoelectronics Engineering, University of Chinese Academy of Sciences, Beijing 100190, China*

[3]*School of Physical Sciences, University of Chinese Academy of Sciences, Beijing 100190, People's Republic of China*

[4]*Ningbo Fengcheng Advanced Energy Materials Research Institute Co., Ltd., Ningbo 315000, China*

[5]*Songshan Lake Materials Laboratory, Dongguan, Guangdong 523808, China*

#These authors contributed equally.

*Corresponding authors:
ygshi@iphy.ac.cn
wzj@iphy.ac.cn

## Abstract

Superconductivity in V-based kagome metals has recently raised great interest as they exhibit the competing ground states associated with the flat bands and topological electronic structures. Here we report the discovery of superconductivity in Ta$_2$V$_{3.1}$Si$_{0.9}$ with a superconducting transition temperature $T_c$ of 7.5 K, much higher than those in previously reported kagome metals at ambient pressure. While the V ions form a two-dimensional breathing kagome structure, the length difference between two different V-V bonds is just 0.04 Å, making it very close to the perfect kagome structure. Our results show that Ta$_2$V$_{3.1}$Si$_{0.9}$ is a moderate-coupled type-II superconductor with a large upper critical field that is close to the Pauli limit. DFT calculations give a van-Hove-singularity band located at Fermi energy, which may explain the relatively high $T_c$ observed in this material.

## Introduction

The kagome lattice, consisting of the layered structure formed by corner-sharing triangles, has garnered extensive attention in condensed matter physics. Kagome antiferromagnetic insulators are one of the ideal platforms to explore the long-sought quantum-spin-liquid states (*1-7*). Kagome metals host intriguing physics as well due to the inherent Dirac cones, van Hove singularities (vHSs), and flat bands at different electron fillings in their electronic structures (*8*). Interestingly, theoretical studies have suggested that unconventional superconductivity can be realized in the kagome lattice within a special range of on-site repulsion and Coulomb interaction at van Hove filling, where the Fermi surfaces are nested and have saddle points on the edges of the Brillouin zone (*9, 10*). However, the emergence of superconductivity is still rare in kagome metals.

A recently discovered family of materials with a V-based kagome lattice, $A$V$_3$Sb$_5$ ($A$ = K, Rb, and Cs) (*11*), exhibits superconductivity with the superconducting transition temperature $T_c$ = 0.93, 0.92, and 2.5 K for K, Rb, and Cs variants, respectively (*12-14*). Moreover, many interesting phenomena have also been discovered, such as anomalous Hall effect (*15*), unconventional chiral CDW order (*16-18*), and topologically nontrivial band structures (*12, 19*). In the superconducting phase of CsV$_3$Sb$_5$, while the Hebel-Slichter coherence peak observed in nuclear magnetic resonance (*20*) and temperature-dependent magnetic penetration depth (*21*) measurements reveal the feature of a conventional *s*-wave superconductor, an unconventional superconducting state has been suggested by other results, such as a finite residual linear term of thermal conductivity at zero magnetic field (*22*), residual zero-energy density of states (DOS) (*23, 24*), and the observation of pair density wave (PDW) (*25*). And these seemingly experimental conflictions may be resolved in a time-reversal-symmetry breaking PDW scenario that arises from the interplay between PDW and a nested Fermi surface (*26*).

Superconductivity also arises in kagome lattice made up of 4*d* or 5*d* elements, such as ternary Laves phase Mg$_2$Ir$_3$Si (*27*) and $R$T$_3$X$_2$ series ($R$ = lanthanide, $T$ = 4*d* or 5*d* transition metal, and $X$ = Si, B or Ga) (*28-34*), where transition metals host an isolated perfect or distorted kagome lattice. In contrast, Sb atoms at the *1a* site of $A$V$_3$Sb$_5$ are located in the center of the V kagome lattice. Therefore, the realization of an isolated V-based kagome lattice is an attractive challenge and may provide intensive insight into the novel phenomena observed in the $A$V$_3$Sb$_5$ family.

In this work, we synthesized a V-based breathing kagome superconductor, Ta$_2$V$_{3.1}$Si$_{0.9}$, with a $T_c$ of 7.5 K. The difference in the side length of the corner-sharing triangles is a mere 2%. A moderate-coupled type-II superconductivity is confirmed by the comprehensive measurement of magnetism, resistivity, and specific heat. Interestingly, the large upper critical field is close to the Pauli limit, which suggests the existence of unconventional behavior. DFT calculations indicate that a vHS band mainly from the V-$d_{yz}$ states is located at Fermi energy, which is strongly relevant to the emergence of the superconductivity.

## Results

**Table 1. Crystallographic data for $Ta_2V_{3.1}Si_{0.9}$ at room temperature.** Space group $P6_3/mmc$ (No. 194), $a$ = 5.0094(8) Å, $c$ = 8.2575(1) Å, V = 179.45(9) Å$^3$.

| atom | x | y | z | site | Occ.[a] | $U_{iso}$[b] |
|------|-----|-----|--------|------|------|--------|
| Ta   | 1/3 | 2/3 | 0.5603(7) | 4f | 1 | 0.01 |
| V1   | 0.1682(2) | 0.3364(5) | 1/4 | 6h | 1 | 0.0077 |
| V2   | 0 | 0 | 0 | 2a | 0.1 | 0.0098 |
| Si   | 0 | 0 | 0 | 2a | 0.9 | 0.0098 |

[a]Th occupancy of each atom. [b]The isotropic displacement parameters are used to avoid an unphysical negative value.

$Ta_2V_{3.1}Si_{0.9}$ crystallizes in a hexagonal $Mg_2Cu_3Si$-type structure, which is also referred to as the ternary C14 Laves phase. The Rietveld refinement of the powder sample, as shown in Fig. 1C, results in the space group $P6_3/mmc$ (No. 194) with cell parameters of $a$ = 5.0094(8) Å and $c$ =8.2575(1) Å, which is consistent with the structure of $Ta_2V_3Si$ reported previously (*35*). A small amount of unknown impurity phase is detected in the powder sample. The atomic coordinates, occupancies, and isotropic displacement parameters are listed in Table 1. The structure of $Ta_2V_{3.1}Si_{0.9}$ consists of $Ta_2Si$ layers and V breathing kagome layers stacking along the crystallographic *c*-axis, as shown in Fig. 1A. The extra V atoms occupy the 2a site of Si atoms. Fig. 1B shows the breathing kagome net of V atoms, in which the shorter and longer distances of the V-V bonds are 2.48Å and 2.53Å, respectively. The difference between the two kinds of V-V bonds is only 2%. The bonds length of the kagome net here is much shorter than those observed in the analogous superconductors, including $KV_3Sb_5$ (2.74Å, V-V bond) (*11*), $LaRu_3Si_2$ (2.84Å, Ru-Ru bond) (*30*), and $Mg_2Ir_3Si$ (2.62 and 2.73Å, Ir-Ir bond) (*27*). Considering a naive perspective that pressure may enhance the critical temperature, the shorter V-V bond in $Ta_2V_{3.1}Si_{0.9}$, seemingly pressing the sample along the kagome plane, implies a higher critical temperature. Moreover, compared with the V-based kagome superconductors $AV_3Sb_5$ (*12-14*), $Ta_2V_{3.1}Si_{0.9}$ exhibits the pure kagome plane without any other atoms, making it highly suitable for studying kagome physics.

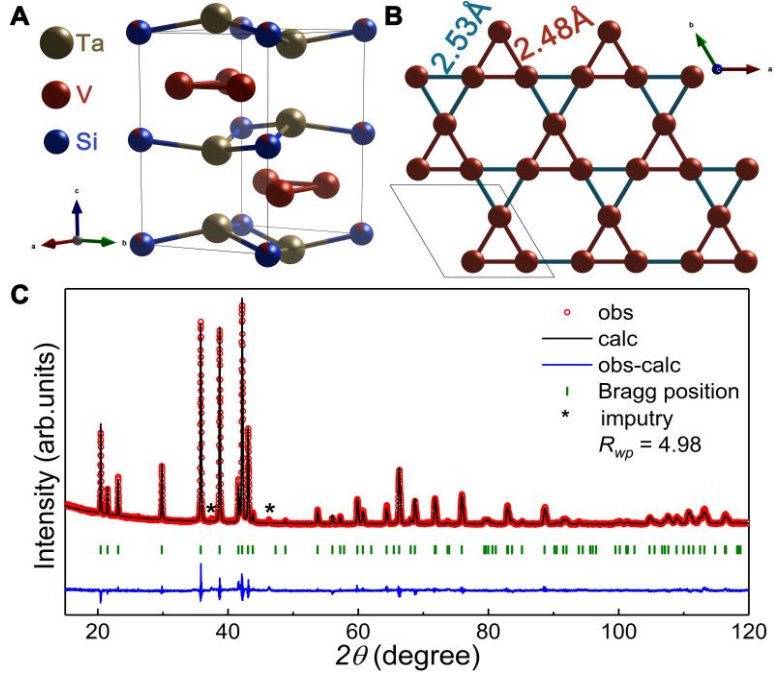

**Fig. 1. Crystal structure of Ta$_2$V$_{3.1}$Si$_{0.9}$.** (**A**) The side view of the crystal structure of Ta$_2$V$_{3.1}$Si$_{0.9}$. (**B**) The breathing kagome lattice of V atoms viewed from the crystallographic *c*-axis. The structure is drawn in the program VESTA (*36*). (**C**) The powder XRD pattern and refined results. The red circles and black lines are observed data and calculated patterns. The green sticks and blue lines represent the Bragg position and the difference between observed and calculated patterns.

The superconductivity of Ta$_2$V$_{3.1}$Si$_{0.9}$ is firstly confirmed by the volume magnetic susceptibility (*4πχ*) measurements under zero-field-cooled (ZFC) and field-cooled (FC) modes from 2 to 10 K with an external magnetic field of 2 mT, as shown in Fig. 2A. The critical temperature (*T$_c$*) of 7.5 K is determined from the intersection between the extrapolated normal state of magnetic susceptibility at low temperature and the line representing the apparent diamagnetic signal (shown by the black lines) (*37, 38*). The superconducting volume fraction reaches almost 100% in the ZFC data, while the value would be lower than 100% considering a demagnetization correction, which can be attributed to the existence of trace impurities detected in the powder XRD patterns. The weak diamagnetic signal observed in the FC process ascribes to the polycrystalline nature of the sample and also the flux pinning effect in a type II superconductor, which is explicitly verified by the loops in isothermal magnetization at 2 K and 6 K, shown in Fig. 2B. Fig. 2C depicts the detailed investigation of the field-dependent magnetization *M(H)* performed at temperatures from 2 K to 8 K with an interval of 1 K. To obtain the lower critical field, $\mu_0H_{c1}$, the low-field data of 2 K are fitted linearly (indicated by the black solid line), and the $\mu_0H_{c1}$ is defined where the data begin to deviate from the fitting line. The extracted $\mu_0H_{c1}$ of each temperature are shown in Fig. 2D and fitted by using the empirical equation:

$$\mu_0H_{c1}(T) = \mu_0H_{c1}(0)\left[1 - \left(\frac{T}{T_c}\right)^2\right]$$

where the fitting parameter $\mu_0H_{c1}(0)$ is the lower critical field at 0 K and $T_c$ is the superconducting transition temperature. The fit yields $\mu_0H_{c1}(0)$ is 3.13(6) mT and $T_c$ is 7.6(1) K, basically coinciding with the critical temperature observed in the $4\pi\chi(T)$ curve.

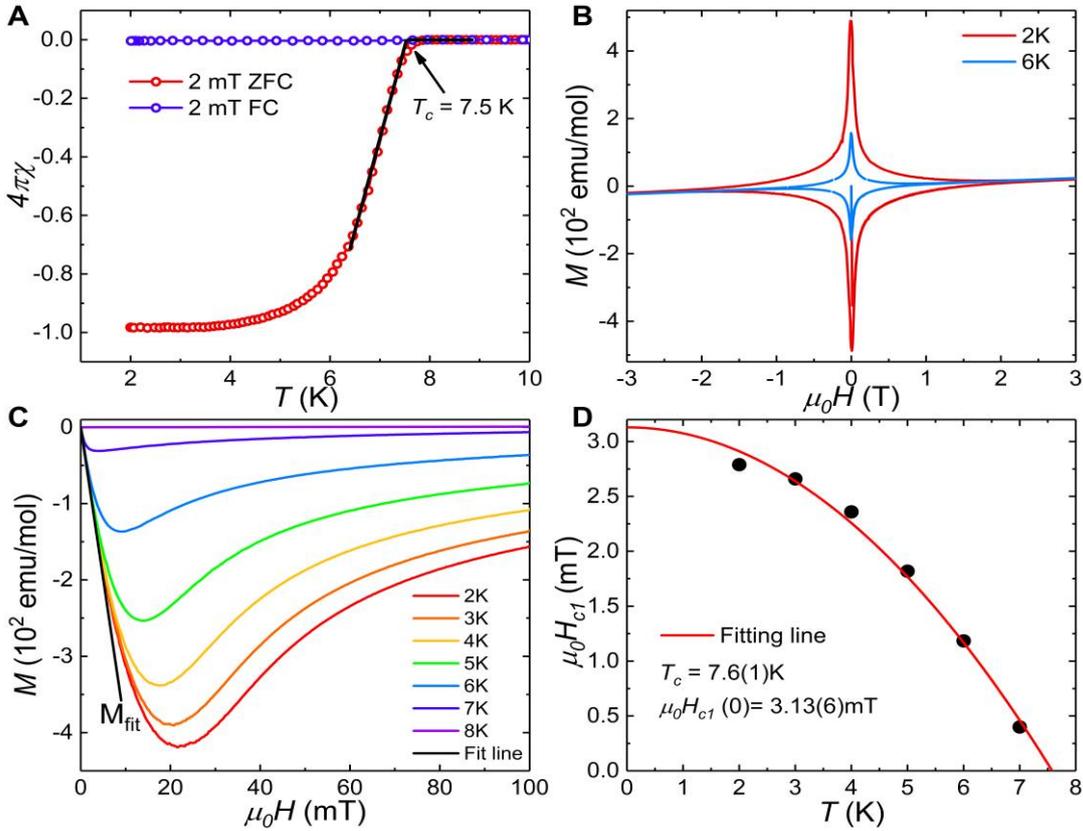

**Fig. 2. The magnetic characterization of $Ta_2V_{3.1}Si_{0.9}$.** (**A**) The temperature-dependent magnetic susceptibility under a magnetic field of 2 mT with ZFC and FC modes. (**B**) The field-dependent magnetization at 2 K and 6 K from -3 T to 3 T. (**C**) The field-dependent magnetization from 2 K to 8 K in an increment of 1 K at low magnetic fields. (**D**) The lower critical field $\mu_0H_{c1}$ versus temperature and the fitting using empirical relation.

The temperature-dependent resistivity $\rho(T)$ measured from 300 K to 2 K is shown in the inset of Fig. 3B. The sample exhibits metallic behavior above the superconducting transition with a large residual resistivity of almost 0.4 mΩ cm. The low-temperature resistivity fitted by the equation $\rho(T) = \rho_0 + AT^n$, where $\rho_0$ is the residual resistivity, $A$ and $n$ are fitting parameters, yields that $\rho_0$ is 0.398 m$\Omega$ cm, $A$ is 1.1(2)*10$^{-6}$, and $n$ is 1.91(4). The value of $n$ close to 2 implies that electron-electron interaction dominates the low-temperature resistivity. Fig. 3A shows $\rho(T)$ from 2 K to 11 K under the increased magnetic field to 15 T with an increment of 1 T. The $T_c$, defined as the temperature where the resistivity drops to 50% of the normal state ($\rho_N$), is 8.1 K. The $T_c$ shifts to the lower temperature with increased fields. Using the aforesaid criteria of $T_c$ for determining the upper critical field ($\mu_0 H_{c2}$), namely the intersection between the $\rho(T)$ curve and the black arrow, we obtain the temperature-dependent $\mu_0 H_{c2}(T)$ in Fig. 3B. The slope near $T_c$ ($H' = -\frac{d\mu_0 H_{c2}}{dT}|_{T_c}$) is 3.05 T/K. This relatively large value gives a high orbital limiting field $\mu_0 H_{c2}^{orb}(0)$ to be 17.12 T, which is determined from $\mu_0 H_{c2}^{orb}(0) = -0.693 H' T_c$ in the dirty limit for a one-band superconductor[39] (which will be discussed later). In the weak-coupled BCS superconductor, the Pauli-limited field $\mu_0 H_P(0)$ is 1.85$T_c$, using the $T_c$ = 8.1 K, the $\mu_0 H_P(0)$ is calculated to be 14.985 T. That the $\mu_0 H_P(0)$ is smaller than $\mu_0 H_{c2}^{orb}(0)$ implies that $\mu_0 H_{c2}$ at low temperature is limited by the Pauli spin susceptibility of the electrons rather than the usual orbital pair-breaking effect, which suggests the anomalous property of the superconductor reported here.

To comprehensively investigate the pair-breaking mechanism in the material, we fit our data based on the WHH model (*39, 40*):

$$\ln\frac{1}{t} = (\frac{1}{2} + \frac{i\lambda_{so}}{4\gamma})\psi(\frac{1}{2} + \frac{\bar{h} + \frac{\lambda_{so}}{2} + i\gamma}{2t}) + (\frac{1}{2} - \frac{i\lambda_{so}}{4\gamma})\psi(\frac{1}{2} + \frac{\bar{h} + \frac{\lambda_{so}}{2} - i\gamma}{2t}) - \psi(\frac{1}{2})$$

where $t = T/T_c$, $\gamma \equiv (\alpha\bar{h})^2 - (\lambda_{so}/2)^{21/2}$, $\psi$ is the digamma function,

$$h^* \equiv \frac{\bar{h}}{(-\frac{d\bar{h}}{dt})_{t=1}} = \frac{\pi^2 \bar{h}}{4} = \frac{H_{c2}}{(-\frac{dH_{c2}}{dt})_{t=1}},$$

$\alpha$ [also known as the Maki parameter (*41, 42*)] and $\lambda_{so}$ are parameters presenting the strength of the spin paramagnetic effect and spin-orbit scattering. Firstly, the data is fitted neglecting the spin paramagnetic effect and spin-orbit scattering ($\alpha$ = 0 and $\lambda_{so}$ = 0), a scenario for the conventional superconductors, while the fitting line (green solid line) deviates the data at low temperatures. Considering the strong Pauli paramagnetic effect in this sample, we obtain the Maki parameter $\alpha$ = 1.62 by substituting the calculated values of $\mu_0 H_{c2}^{orb}(0)$ and $\mu_0 H_P(0)$ into the formula (*42*):

$$\alpha = \frac{\sqrt{2} H_{c2}^{orb}(0)}{H_p(0)}$$

By fixing the value of $\alpha = 1.62$ and adjusting the $\lambda_{so}$, the data can be well-fitted by the WHH model (red solid line), yielding the $\mu_0 H_{c2}(0) = 14.2$ T. Noticing that the parameter $\lambda_{so} = 2.2$ is essential to depict the experiment data, and the fitting line ignoring the $\lambda_{so}$ ($\alpha = 1.62$ and $\lambda_{so} = 0$) is also shown with green dashed line to give a vivid contrast. This suggests that the spin paramagnetic effect and spin-orbit scattering are important to describe the upper critical field for the material. The appearances of large $\mu_0 H_{c2}(0)$ and deviation of the WHH model are also observed in several cases: non-centrosymmetric superconductors (*37*), heavy-fermion superconductors (*43, 44*), iron-based high-temperature superconductors (*45, 51*), and deficiencies-induced strong spin-orbit scattering system (*52*). The first three scenarios can be easily excluded from our material, and we argue that spin-orbit scattering plays an important role in enhancing the upper critical field for the reason of the large value of $\lambda_{so}$ here and polycrystalline samples suffering in dirty limit.

The obtained $\mu_0 H_{c2}(0)$ is used to determine the Ginzburg-Landau coherence length $\xi_{GL}$ from the following relation:

$$\mu_0 H_{c2}(0) = \frac{\Phi_0}{2\pi \xi_{GL}^2(0)}$$

where $\Phi_0 = h/2e$ is the magnetic flux quantum. This leads to the $\xi_{GL}$ 48.14 Å, a distinctly shorter coherence length than other kagome superconductors, such as LaRu$_3$Si$_2$ ~ 107 Å (*53*), LaIr$_3$Ga$_2$ ~ 85 Å (*33*), and Mg$_2$Ir$_3$Si ~ 74 Å (*27*). Employing the results of $\xi_{GL}$ and $\mu_0 H_{c1}(0)$ calculated previously, the magnetic penetration depth $\lambda_{GL} = 4960.45$ Å is estimated using the following equation:

$$\mu_0 H_{c1} = \frac{\Phi_0}{4\pi \lambda_{GL}^2} \ln \frac{\lambda_{GL}}{\xi_{GL}}$$

Then the Ginzburg-Landau parameter $\kappa_{GL} = \lambda_{GL}/\xi_{GL} = 103.03 > 1/\sqrt{2}$, confirms that the material is a type-II superconductor. Using the result of $\mu_0 H_{c1}(0)$, $\mu_0 H_{c2}(0)$, and $\kappa_{GL}$, the thermodynamic critical field $H_c$ is determined from the relation:

$$H_{c1} H_{c2} = H_c^2 \ln \kappa_{GL}$$

which yields $\mu_0 H_c(0) = 97.45$ mT.

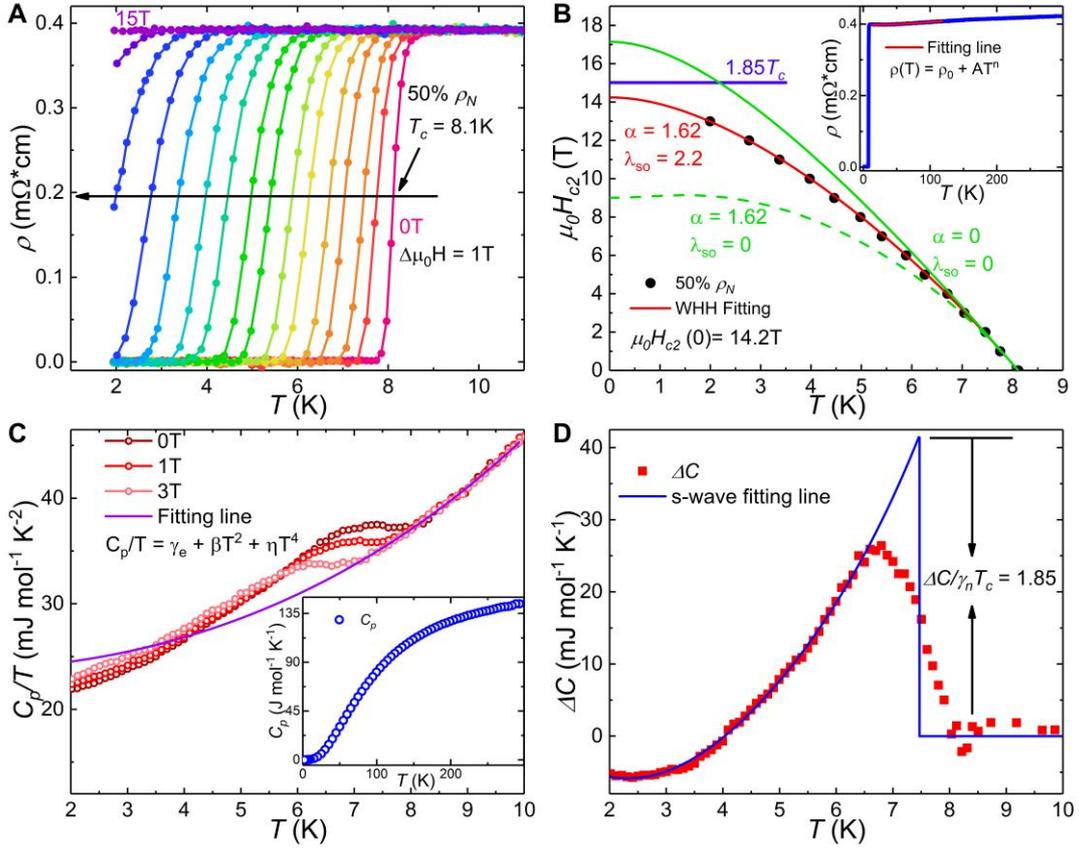

**Fig. 3. The resistivity and specific heat of $Ta_2V_{3.1}Si_{0.9}$.** (**A**) The temperature-dependent resistivity under magnetic field from 0 T to 15 T in an increment of 1 T. (**B**) The upper critical field versus temperature data and the WHH fitting. The inset shows the temperature dependence of resistivity from 300 K to 2 K. (**C**) The temperature-dependent specific heat under various magnetic fields. The inset illustrates the specific heat from 300 K to 2 K. (**D**) $\Delta C$ versus $T$ data from 10 K to 2 K. The blue line below the critical temperature is the *s*-wave fitting.

The temperature-dependence of specific heat $C_p(T)$ measurement is performed to finally confirmed the bulk superconducting nature of the material. No anomaly, such as CDW transition in $AV_3Sb_5$ (A = K, Rb, Cs) (*11*), is observed above the superconducting transition temperature, as shown in the inset of Fig. 3C. Low-temperature data with applied field from 0 T to 3 T exhibits obvious anomaly that shifts to lower temperatures with increased fields, corresponding to the emergence of superconducting transition, as shown in Fig.3C. The $C_p(T)$ data at 3 T is described (purple solid line) by the following relation:

$$\frac{C_p}{T} = \gamma_e + \beta T^2 + \eta T^4$$

where $\gamma_e T$ is the electronic contribution and $\beta T^3 + \eta T^5$ is the phonon contribution to the specific heat. The fitting yields $\gamma_e$ = 23.7(9) mJ mol$^{-1}$ K$^{-2}$, $\beta$= 0.018(2) mJ mol$^{-1}$ K$^{-4}$, and $\eta$ = 0.0003(1) mJ mol$^{-1}$ K$^{-6}$. The Debye temperature $\Theta_D$ can be calculated using the value of β and the equation:

$$\Theta_D = (\frac{12\pi^4}{5\beta} nR)^{1/3}$$

which gives the $\Theta_D$ 409.2(5) K. Combining the $\Theta_D$ = 409.2 K and $T_c$ = 7.5 K (determined from $4\pi\chi$ (T) curve), the electron-phonon coupling strength $\lambda_{ep}$ can be calculated by using the McMillian equation (54):

$$\lambda_{ep} = \frac{1.04 + \mu^*\ln(\frac{\Theta_D}{1.45T_c})}{(1-0.62\mu^*)\ln\left(\frac{\Theta_D}{1.45T_c}\right) - 1.04}$$

where $\mu^*$ is a typical value of 0.13. The calculated value of $\lambda_{ep}$ is 0.66, suggesting a moderate-coupled superconductor. The density of electronic states at the Fermi energy $N(E_F)$ can be obtained from the following relation (37):

$$N(E_F) = \frac{3\gamma_e}{\pi^2 k_B^2 (1+\lambda_{ep})}$$

Where $k_B$ is the Boltzmann constant. Employing the obtained values of $\gamma_e$ = 23.7 mJ mol$^{-1}$ K$^{-2}$ and $\lambda_{ep}$ = 0.66, the $N(E_F)$ is estimated to be 5.89 states eV$^{-1}$ per formula unit (f.u.). The relatively large $N(E_F)$ is consistent with the existence of a van-Hove-singularity band near the Fermi energy (discussed in the electronic structure part). The mean free path $l$ is estimated from the relation (55):

$$l = 2.732 \times 10^{-14} \frac{(\frac{m^*}{m_e})V_M^2}{N(E_F)^2 \rho_0}$$

where $V_M$, $m^*$, and $m_e$ are the molar volume, the effective mass of the individual quasiparticles, and free-electron mass, respectively. Inserting $\left(\frac{m^*}{m_e}\right) = 1$ and the obtained $N(E_F)$ and $\rho_0$ into the above expression gives $l$ = 0.5 Å, which is far less than $\xi_{GL}$ = 48.14 Å. Therefore, the sample is within the dirty limit.

To inspect the pairing symmetry and the magnitude of the specific-heat jump corresponding to the superconducting state $\Delta C$, we obtain $\Delta C$ by subtracting the normal state value (the fitting line) from the zero-field data C(0 T), $\Delta C$ = C(0 T) - $C_{fit}$ = $C_{es}$ - $\gamma_n T$, as shown in Fig. 3D, where $C_{es}$ is the superconducting quasiparticle contribution and the $\gamma_n$ presents the normal state Sommerfeld coefficient of the superconducting part (38, 56). The entropy of the superconducting state $S_{es}$ is expressed by the following equation:

$$S_{es} = -\frac{3\gamma_n}{k_B \pi^3} \int_0^{2\pi} \int_0^\infty [(1-f)\ln(1-f) + f\ln f] \, d\varepsilon d\phi$$

where $f$ is the quasiparticle occupation function $f = (1 + e^{E/k_B T})^{-1}$ and $E = \sqrt{\varepsilon^2 + \Delta^2(\phi)}$. $\Delta(\phi) = \alpha \Delta_{BCS}(T)$ is the angle-independent gap function for an *s*-wave superconductor. Here $\Delta_{BCS}(T)$ is the weak-coupled BCS gap function. The electronic specific heat is calculated by $C_{es} = T(\partial S_{es}/\partial T)$. The experimental data are fitted by the *s*-wave model, resulting in $\alpha$ =1.12, $\gamma_n$ =3.1 mJ mol$^{-1}$ K$^{-2}$, and $T_c$ =7.47 K. The normalized specific heat jump $\Delta C/\gamma_n T_c$ is obtained to be 1.85, a larger value than the expected value of 1.43 for a weak-coupled BCS superconductor. Meanwhile, the value of $2\Delta(0)/k_B T_c$ is calculated to be 3.93, which is also larger than the BCS theory value of 3.52. A summary of all the obtained superconducting parameters is in Table 2.

Table 2. Superconductivity parameters of Ta$_2$V$_{3.1}$Si$_{0.9}$.

| Parameter | Units | Ta$_2$V$_{3.1}$Si$_{0.9}$ |
|---|---|---|
| $T_c$ | K | 7.5 (from $4\pi\chi$-$T$) |
| $\mu_0 H_{c1}(0)$ | mT | 3.13(6) |
| $\mu_0 H_{c2}(0)$ | T | 14.2 |
| $\mu_0 H_c(0)$ | mT | 97.45 |
| $\mu_0 H_p(0)$ | T | 14.985 |
| $\xi_{GL}$ | Å | 48.14 |
| $\lambda_{GL}$ | Å | 4960.45 |
| $\kappa_{GL}$ | -- | 103.03 |
| $l$ | Å | 0.50 |
| $\gamma_e$ | mJ mol$^{-1}$ K$^{-2}$ | 23.7(9) |
| $\Delta C/\gamma_n T_c$ | -- | 1.85 |
| $\lambda_{ep}$ | -- | 0.66 |
| $N(E_F)$ | state eV$^{-1}$ per f.u. | 5.89 |
| $\Theta_D$ | K | 409.2 |
| $\Delta(0)$ | meV | 1.27 |
| $2\Delta(0)/k_B T_c$ | -- | 3.93 |

To better understand the properties of Ta$_2$V$_{3.1}$Si$_{0.9}$, we performed the DFT calculations for the pristine (Ta$_2$V$_3$Si) and V-doped (Ta$_2$V$_{3.1}$Si$_{0.9}$) crystals. Their band structures are presented in Fig. 4A and 4B, respectively. The density of states (DOS) of Ta$_2$V$_{3.1}$Si$_{0.9}$ is shown in Fig. 4B as well. In the orbital-weighted band structure of Fig. 4A, one can find a vHS band is mainly from the V-$d_{yz}$ states, which is about 200 meV above the Fermi energy ($E_F$). In the Ta$_2$V$_{3.1}$Si$_{0.9}$, this vHS band significantly shifts downwards and is located at $E_F$, highlighted with a green stripe. To elucidate the critical role of the vHS in the emergence of superconductivity, we conduct comprehensive low-temperature measurements of Ta$_2$V$_3$Si, as shown in Fig. S1 to S2. Although the magnetic susceptibility and resistivity show superconductivity at 5.1 K and 6.0 K, respectively, such low superconducting volume fraction and the absence of obvious specific heat jump meticulously demonstrate that the Ta$_2$V$_3$Si is not a bulk superconductor. Moreover, the *d* states from V atoms are predominant in the DOS, indicating the significant role of *d* electrons of the V breathing kagome lattice for the electronic properties of Ta$_2$V$_{3.1}$Si$_{0.9}$. The calculated DOS($E_F$) value is 4.88 states eV$^{-1}$ per f.u., which generally coincides with the value presented in Table 2. The constant energy surfaces of the band at $E_F$ are plotted in the topside of Fig. 4C; the bottom shows iso-energy surfaces at $E = E_F - 0.118$ eV, at which the vHS of M point of Brillouin zone boundary locates.

The superconducting transition temperature is estimated using Allen-Dynes modified McMillian equation:

$$T_c = \frac{\omega_{log}}{1.2\ k_B} \exp\left[\frac{-1.04(1+\lambda)}{\lambda(1-0.62\mu^*)-\mu^*}\right]$$

where $k_B$ is the Boltzmann constant, $\mu^*$ is the effective screened Coulomb repulsion constant. $\lambda = \sum_{qv} \lambda_{qv} = 2\int_0^\infty d\omega \frac{\alpha^2 F(\omega)}{\omega}$ is the electron-phonon coupling constant, and $\omega_{log}$ is the logarithmic average phonon frequency. With $\mu^* = 0.1$ and $\lambda = 0.5794$, $T_c$ is estimated to be 4.87 K, which is slightly smaller than the experimental value. The phonon spectrum, Eliashberg spectral functions $\alpha^2F(\omega)$, and the frequency-dependent coupling $\lambda(\omega)$ of Ta$_2$V$_{3.1}$Si$_{0.9}$ is shown in Fig. 4D. The contributions of $\lambda$ mainly come from the phonon modes of 180 cm$^{-1}$ < $\omega$ < 220 cm$^{-1}$. Among these phonon modes, it can be found that the 18$^{th}$ and 19$^{th}$ phonon bands at M point have higher $\lambda_{qv}$ than others. We have selected the phonon mode of the 19$^{th}$ phonon bands at M point (B$_{1g}$ mode) and plotted its top view and side view, as it is shown in Fig. 4E, which depicts that the in-plane vibration mode of V atoms mainly contributes.

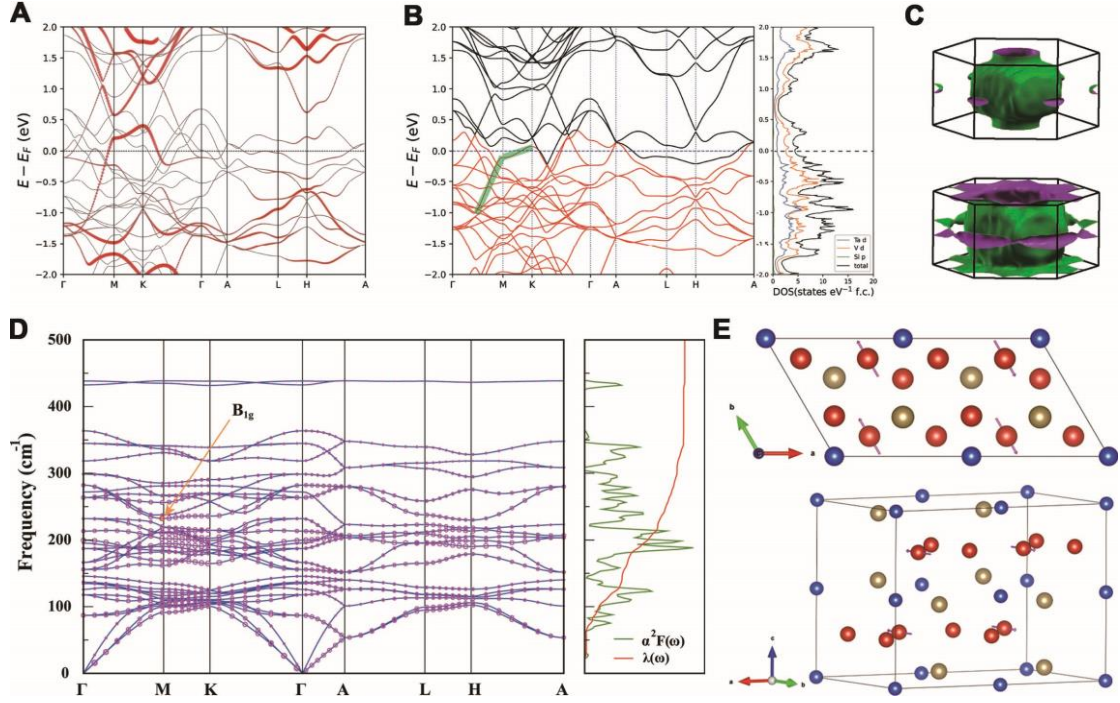

**Fig. 4. The DFT calculation of Ta$_2$V$_3$Si and Ta$_2$V$_{3.1}$Si$_{0.9}$.** (**A**) The orbital-resolved band structure (Ta$_2$V$_3$Si) of V atoms' local d$_{yz}$ orbitals. (**B**) The band structure and density of states of Ta$_2$V$_{3.1}$Si$_{0.9}$. The highlight band (29$^{th}$ band at M-K) drops to $E_F$ after being doped. (**C**) The iso-energy surface at $E = E_F$ (above) and $E = E_F$ - 0.118 eV (below) of the 29$^{th}$ band of Ta$_2$V$_{3.1}$Si$_{0.9}$. (**D**) The phonon spectrum, Eliashberg spectral functions $\alpha^2F(\omega)$, and the frequency-dependent coupling $\lambda(\omega)$ of Ta$_2$V$_{3.1}$Si$_{0.9}$. The electron-phonon couplings $\lambda_{q\nu}$ are depicted by magenta circles. (**E**) An illustration depicting the top and side views of the B$_{1g}$ phonon bands' vibrations, with arrows indicating the direction of the atomic motion.

## Discussion

In summary, we investigate the structure and superconductivity of Ta$_2$V$_{3.1}$Si$_{0.9}$. The material crystallizes in the layered hexagonal structure with an isolated breathing kagome plane of vanadium atoms. Comprehensive measurements of magnetism, resistivity, and specific heat demonstrate that Ta$_2$V$_{3.1}$Si$_{0.9}$ is a moderate-coupled superconductor with a critical temperature of 7.5 K. Although most of the physical properties are close to the BCS theory, the large upper critical field, distinct spin paramagnetic effect, and spin-orbit scattering suggest an unusual pairing mechanism. Moreover, DFT calculations depict a vHS band deriving from the V-d$_{yz}$ states locates at $E_F$, which is also observed in CsV$_3$Sb$_5$ and plays an important role in the formations of superconductivity and CDW order (*57, 58*). Our study provides a new platform to research the interplay between geometrical frustration and superconductivity in the V-based kagome lattice. To further explore the novel phenomena in this material, such as nontrivial topological band structure, CDW order, time-reversal symmetry breaking, and PDW observed in other V-based kagome superconductor $A$V$_3$Sb$_5$, the single crystals of Ta$_2$V$_{3.1}$Si$_{0.9}$ are the urgent need.

## Materials and Methods

**Synthesis.** Polycrystalline samples of Ta$_2$V$_{3.1}$Si$_{0.9}$ were synthesized by using the arc-melting method. High-quality ingredients Ta (5.913g), V (2.58g), and Si (0.413g) were weighed out in a molar ratio of 2: 3.1: 0.9, arc-melted thrice under argon an atmosphere, utilizing a titanium ingot as an oxygen getter. The sample was then annealed in vacuumed quartz tube at the temperature of 800°C for 3 days. The polycrystalline ingot was silvery and stable in the air. Rectangular-shaped samples were obtained for physical property measurements by using the diamond wire-cutting machine.

**Crystal Structure.** The powder X-ray diffraction (PXRD) on crushed samples was performed in Rigaku SmartLab 9kW with Cu K$\alpha$ radiation. The crystal structure was refined with the program package GSAS-II suit (*59*).

**Physical Property Measurement.** Magnetic susceptibility $\chi(T)=M/H$ and isothermal magnetization $M(H)$ were measured using a Physical Property Measurement System (PPMS Dynacool, Quantum Design) equipped with a vibrating sample magnetometer (VSM) option. A sample with a weight of 14 mg was chosen for heat capacity measurement in the same PPMS. The standard four-probe method was employed for electric measurement in PPMS-16T.

**Electronic Structure Calculations.** Calculations of the density functional theory were performed using the Vienna ab initio package (VASP) and quantum espresso (QE) with projector augmented wave (PAW) method and Perdew-Burke-Ernzerhof (PBE) exchange-correlation functional. A plane wave energy cutoff of 420 eV and a 16×16×10 $k$-mesh were employed. We adopted the virtual crystal approximation (VCA) to simulate doping cases.

## References

1. P. W. Anderson, Resonating valence bonds: A new kind of insulator? *Mater. Res.*

## Acknowledgments

We would like to thank Yi Zhou for the helpful discussion and careful reading of the manuscript.

**Funding:** This work was financially supported by the National Key Research and Development Program of China (2022YFA1403400)

The Natural Science Foundation of China (Grant No. U2032204, U22A6005, and 12104492)

The Strategic Priority Research Program of the Chinese Academy of Sciences (XDB33010000)

The China Postdoctoral Science Foundation (Grant No. 2021TQ0356),

The Informatization Plan of Chinese Academy of Sciences (CAS-WX2021SF-0102)

The K. C. Wong Education Foundation (Grants No. GJTD-2020-01)

The Synergetic Extreme Condition User Facility (SECUF).

**Author contributions:** H.X.L. grew the samples and measured the transport data. H.X.L., J.M.S., Y.L., and D.Y.Y. carried out the analysis. J.Y.Y. and Z.L.Y. carried out DFT analyses and provided theoretical support. D.H.C., H.L.F., S.L.L., Z.J.W., and Y.G.S. are the principal investigators. **Competing interests:** The authors declare that they have no competing interests. **Data and materials availability:** All data are available in the main text or the supplementary materials.

## Supplementary Materials

Crystallographic information file of $Ta_2V_{3.1}Si_{0.9}$.

Fig. S1: XRD patterns of $Ta_2V_3Si$ and $Ta_2V_3Si_{0.9}$.

Fig. S2: Low-temperature magnetic susceptibility, resistivity, and specific heat measurements of Ta$_2$V$_3$Si.

# Supplementary Materials for

**Vanadium-Based Superconductivity in a Breathing Kagome Compound Ta$_2$V$_{3.1}$Si$_{0.9}$**

HongXiong Liu *et al.*

# XRD patterns of Ta$_2$V$_3$Si and Ta$_2$V$_{3.1}$Si$_{0.9}$

Fig. S1 shows XRD patterns of Ta$_2$V$_3$Si and Ta$_2$V$_3$Si$_{0.9}$. One can find that there are more impurity phases in the Ta$_2$V$_3$Si sample, as shown in the blue areas.

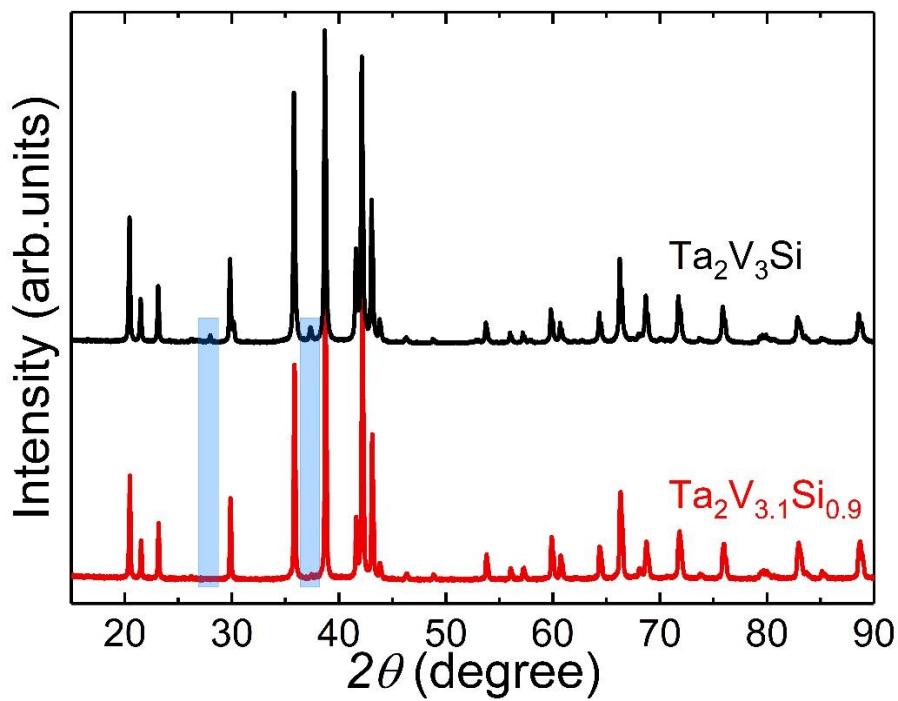

**Fig. S1. XRD patterns of Ta$_2$V$_3$Si and Ta$_2$V$_3$Si$_{0.9}$.**

# Low-temperature physical properties of Ta$_2$V$_3$Si

Fig. S2A shows the volume magnetic susceptibility from 2 to 10 K with a magnetic field of 2 mT. The ZFC data exhibit a diamagnetic signal below the temperature of 5.1 K. However, the superconducting volume fraction only reaches 50% before subtracting the demagnetization factor, which implies that this sample may be not a bulk superconductor but contains some superconducting grain boundary. Fig. S2B shows $\rho(T)$ from 2 K to 10 K under the increased magnetic field to 9 T with an increment of 1 T. The $T_c$ obtained here is only 6.0 K and decreases to 2.3 K under the magnetic field of 9 T, which is rather lower than observed in Ta$_2$V$_{3.1}$Si$_{0.9}$. The absence of an obvious specific heat jump again proves the hypothesis that the stoichiometric Ta$_2$V$_3$Si is not a bulk superconductor, as shown in Fig. S2C.

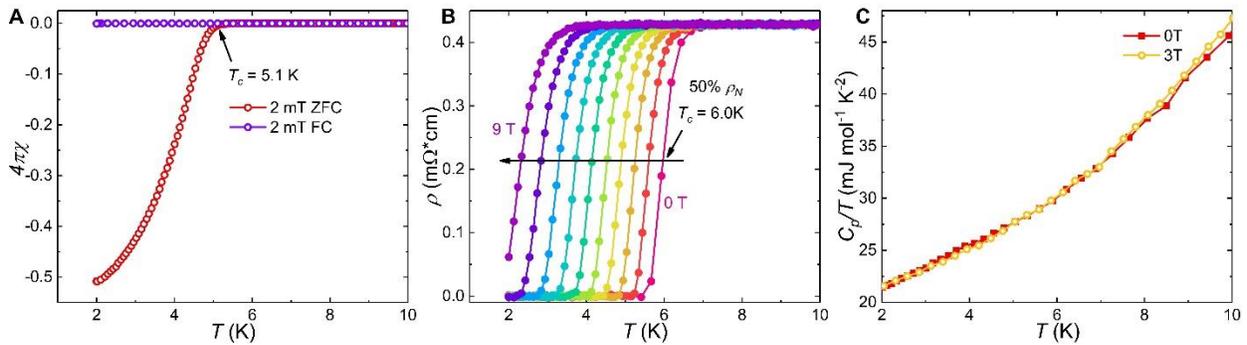

**Fig. S2. Low-temperature measurements of Ta$_2$V$_3$Si.** (**A**) The temperature-dependent magnetic susceptibility under a magnetic field of 2 mT with ZFC and FC modes. (**B**) The temperature-dependent resistivity under magnetic fields from 0 T to 9 T in an increment of 1 T. (**C**) The temperature-dependent specific heat under magnetic fields 0 T and 3 T.